\documentstyle[12pt]{article}
\newlength{\lll}
\newlength{\lla}
\newcount\eqnum
\def\eqn#1{\global\advance\eqnum by 1
   \xdef#1{ (\the\eqnum ) }(\the\eqnum )  }
\begin{document}
\title{%
Noncomputability Arising In  Dynamical Triangulation Model Of Four-Dimensional
Quantum Gravity}
\author{A. Nabutovsky$^1$\\
 { Department of Mathematics,} \\
  {University of Toronto, Toronto, Ontario, M5S 1A1, Canada}\\
and\\
R. Ben-Av$^2$\\
{Physics Department,} \\
{Princeton University, Princeton, N.J. 08540,USA}}
\maketitle

\begin{abstract}

Computations in Dynamical Triangulation Models of Four-Dimensional Quantum
Gravity involve weighted averaging over sets of all distinct
triangulations of compact four-dimensional manifolds. In order to be able to
perform such computations one needs an algorithm which for any given $N$
and a given compact four-dimensional manifold $M$
constructs all possible triangulations of $M$ with $\leq N$ simplices. Our
first result is that such algorithm does not exist. Then we discuss
recursion-theoretic limitations of any algorithm designed to perform
approximate calculations of sums over all possible triangulations of a
compact four-dimensional manifold.

\end{abstract}
\vskip 2cm
\begin{flushright}
PUPT-1327; Version 1.2.\\
BITNET: 1- alex@math.toronto.edu ; 2 - radi@acm.princeton.edu
\end{flushright}
\newpage
A well-known problem in physics is to unify Quantum Theory with General
Relativity. One of the proposed approaches 
involves integration over the space of metrics on all compact four
dimensional manifolds. The integration is done separately for any
particular topological type of the four dimensional manifolds.
Then one would like to sum over all topological types attributing
an appropriate weight to every topological type.

Although there is no mathematically rigorous definition of a
measure with the required properties
on the (infinite dimensional) space of all metrics on a
four dimensional manifold (but see [Po] for the two
dimensional case), recently the following idea to do this computation
was proposed. One considers a kind of grid in the space of all
metrics. This grid is formed by metrics which are defined as
follows: One starts from a triangulation of the smooth four
dimensional manifold of interest. Then one considers all possible
 triangulations of the manifold combinatorially equivalent
to the chosen initial triangulation.
One does not distinguish between simplicially isomorphic triangulations.
(From now on we will mean by a {\it triangulation} of a PL-manifold $M$ a
simplicial complex $K$  such that the corresponding polyhedron $\vert K\vert$
is PL (piecewise-linearly) homeomorphic to $M$. This definition of
triangulations somewhat differs from the standard one but is more
convenient for our aims.  Usually one requires only
that $\vert K\vert$ be homeomorphic to $M$. We will not distinguish between
simplicially isomorphic triangulations.)
For any of these triangulations one
assigns the unit length for any of its 1-dimensional
simplices. In this way one gets for any triangulation a
distance function on the manifold. The metrics corresponding to
all possible triangulations are considered as a uniform grid in
the space of all metrics. Afterwards one can approximate the
necessary integrals by means of sums over all the nodes of the
grid (see [BKKM] ,[BD],[M] ,[ADF] for the two dimensional case
and [AM] ,[V] ,[AJ] for the
four dimensional case).
To generate this grid, one uses the following method: One starts
from a prescribed point (i.e. a given triangulation of the
manifold). Then one makes elementary operations in order to move from a
point to its neighboring
points. (More precisely, one introduces a finite set of elementary operations.
Operations from this set change any particular
triangulation
to some other triangulations which by definition are considered
the neighboring triangulations of this particular triangulation).

In practice the whole grid is not generated, but rather a
probabilistic approximation is introduced. A Markov chain is
defined, and the necessary sum is computed by means of a Monte-
Carlo method. In order to validate this procedure one needs to
fulfill in particular the following requirement of {\it ergodicity}:
Using the set of elementary moves one should be able to get any
triangulation from any other triangulation. \par
{}From the computational
point of view it is more sensible to impose the following
stronger constraint of {\it computational ergodicity} on the
considered set of elementary moves:
There exists a recursive function $r$ such that for any $N$ and any two
prescribed triangulations with $\leq N$ simplices
there exists a sequence of not more than $r(N)$ elementary moves
which transforms one triangulation to the other.\par
The first main point of this note is the observation that in
the dynamical triangulation model for four
dimensional quantum gravity
there is no finite set of elementary operations such that
the {\it computational ergodicity} will hold (although,
as it was shown in [GV], the set of
elementary moves considered in [AM], [V], [AJ] satisfies
the ergodicity requirement).
That is, for every finite
set of elementary operations and for every recursive function
$r(N)$, there always exist some $N$ and
two triangulations with $N$ simplices
such that the number of operations needed to transform one
triangulation to the other exceeds $r(N)$.

This observation is based on the classical result of Markov on
algorithmic unrecognizability of a specific four dimensional
manifold (cf. [BHP]).
(As it follows from the proof of this result given in
[F], Theorem 14.1, this manifold can be taken, for example,
 diffeomorphic to the four dimensional sphere with 46 attached
handles of index two. Denote this manifold by $S_0$.)
The proof of this result of Markov
is based on results of Rabin and Adyan on algorithmic
unsolvability of the triviality problem for
finitely presented groups.
More precisely, Markov proved that there is no
algorithm which for a given, finitely presented, four dimensional manifold
verifies whether or not it is homeomorphic (or diffeomorphic, or PL-
homeomorphic) to $S_0$.
Furthermore we would like to mention the related
result of S. P. Novikov (published as ch. 10 of [VKF]),
which states the algorithmic unrecognizability of spheres
$S^n$ for any $n\geq 5$. Although it is not known now whether or
not the four dimensional sphere can be recognized in the class of
all PL (or smooth) four dimensional manifolds, it seems very plausible
that the four dimensional sphere is also algorithmically
unrecognizable. In all these unrecognizability results for PL-manifolds
all PL-manifolds can be assumed presented by some triangulation.\par

{\bf Proposition 1.} {\it Let $M_0$ be a PL-manifold that can not be recognized
by any algorithm. Then there is no
finite set of elementary moves on the set of all triangulations of $M_0$
satisfying the requirement of computational
ergodicity.}\par\par

{\bf Proof.}  One can prove this proposition by contradiction using an argument
very similar to the argument used in [ABB] to prove
Theorem 5 b) there. Indeed, suppose that a finite set of elementary
moves on the set of all triangulations of $M_0$ satisfying the requirement of
computational ergodicity exists. Then, applying all possible chains of the
moves to any fixed given triangulation of $M_0$
one can generate the list of all possible triangulations of $M_0$ with $\leq
 N$ simplices in time recursively depending on $N$. Hence given
a manifold $M$ presented by some its
triangulation $T$ with $N(T)$ simplices one can
decide whether or not $M$ is PL-homeomorphic to $M_0$ as follows: First,
it is necessary to generate the list of all triangulations of $M_0$ with
$\leq N(T)$ simplices. Then, for any triangulation on this list one checks
whether or not it is simplicially
isomorphic to $T$. (This step evidently can be
effectively done; cf. [ABB, Lemma 2.16].)
 Hence, the assumption that there exists a computationally
ergodic finite set of elementary moves on the set of
triangulations of $M_0$ implies
the existence of an algorithm checking for any manifold $M$ presented by
some triangulation whether or not $M$ is PL-homeomorphic to $M_0$. Q.E.D.
\medskip \par
Thus, at least for the mentioned manifold $S_0$
there is no computationally ergodic finite set of elementary moves
on the set of its triangulations.
Moreover,
if $S^4$ cannnot be effectively recognized in the class of all 4-dimensional
manifolds, then
there is no
computationally ergodic finite set of elementary moves on the set of
triangulations of $S^4$.

Hence, in general, it is natural to expect that a computation
of integrals within to accuracy $\epsilon$ using the described above approach
will require an amount of time growing
non-recursively fast with $[1/\epsilon]$. (Of course, if the integrated
function
is of a special form this can be not the case. ) \par
\medskip
Consider now an ergodic finite set of elementary moves on the space of
triangulations of
a compact four-dimensional manifold $M_0$ which cannot
be recognized by any algorithm. Let $T_0$ be an arbitrary triangulation of
$M_0$. Proposition 1 immediately implies
that for any recursive function $t(N)$ there exist arbitrary large $N$
such that some triangulations of $M_0$ with $N$ simplices cannot be
obtained from $T_0$ by less than $t(N)$ elementary moves. Taking, for
example, $t(N)= [\exp(\exp(\ldots\exp(N)))]$ (the exponentiation is performed
$N$ times), we see that from the practical computational point of
view some part of the grid will be out of reach for large $N$. Now the
following
question naturally arises:\par
 Which part of the grid will be out of reach asymptotically , when $N$ tends
   to infinity? More precisely, let $A(N)$ be an algorithm
 which for any $N$ produces some amount of distinct
triangulations of $M_0$ with $\leq N$ simplices. Denote the number of
triangulations with $\leq N$ simplices produced by $A$ by $s_A(N)$.
Denote by $s(N)$ the total number
of triangulations of $M_0$ with $\leq N$ simplices. Now our problem can be
formulated as determining
the value of $S$ defined by the following formula:
\begin{equation}
S\ =\ \sup_A\lim\ \sup_{N\longrightarrow\infty}s_A(N)/s(N).
\end{equation}
(The maximum is taken over the set of all possible algorithms.)\par
First, note that $s_A(N)$ is a recursive function. Now let us prove the
following proposition:\par
{\bf Proposition 2.} {\it $s(N)$ is a non-recursive function.}\par\par
{\bf Proof:} In order to see that $s(N)$ is not recursive assume
the opposite. Then we
have the following algorithm recognizing for a given polyhedron $P$
whether or not it is PL-homeomorphic to $M_0$ (providing $M_0$
and $P$ are
presented by some triangulations $T_0$ of $M$ and $T$ of $P$ ):\\
 Put $N$ to be equal to the number of simplices in $T$.  Compute $s(N)$.
Apply successively
all possible finite combinations of the Alexander move
and its inverse
(see [A],[GV]) to $T_0$,
keeping record of the number of distinct triangulations of $M_0$,
which are already obtained. If this number is equal to $s(N)$, then stop.
Since the Alexander move and its inverse form an ergodic set of moves on
the set of triangulations of any compact manifold ([A]), eventually we shall
obtain the list of all distinct triangulations of $M_0$ with $\leq N$
simplices.
The last step of the algorithm will be
to compare $T$  with all $s(N)$ triangulations from the obtained list of
triangulations of $M_0$ with $N$ simplices. The existence of this
algorithm provides the desired contradiction which proves
the non-recursiveness of $s(N)$. Q.E.D.\par\par

(This non-recursiveness result has the following direct physical
interpretation.
The partition function defined in [AM] for Quantum Gravity is :
\begin{equation}
{\cal Z}(\hat \lambda_4, \hat \lambda_0)  = \sum_{\rm triangulations \
of \ a \ manifold }
\exp \left ( -\hat \lambda_4 N_4 - \hat \lambda_0
N_0 - \Delta\lambda_4(N_4 - \hat N_4)^2 - \Delta\lambda_0(R-\hat R)^2
\right ).
\end{equation}
Where $N_0,N_1,N_2,N_3,N_4$ are the numbers of $0,1,2,3,4$-simplexes in the
simplicial complex respectively, $\hat \lambda_4$,$\hat \lambda_0$,
$\Delta\lambda_4$,
$\Delta\lambda_0$,$\hat N_4$ and $\hat R$ are the parameters of the model and
 $R$ is defined by :
\begin{equation}
R= 4 \pi(N_0 + N_4 -2) -10\alpha N_4 \ \ , \ \ \
\alpha = {\rm arccos ({\frac{1}{Dimension}}}).
\end{equation}
This partition function can be defined for any topological type of
compact smooth four-dimensional manifolds. In [AM] it was considered for
$S^4$. Consider it for the manifold $S_0$.
Note that $s(N)$ equals the partition function defined above, where
$\hat\lambda_0=\hat\lambda_4=\Delta\lambda_0 = 0$ and $\Delta\lambda_4
= \infty $,$\hat N_4 = N$.
As an easy corollary one can see that $\Delta\lambda_4=\infty$ is a sufficient
condition for the partition function $Z(\hat\lambda_4,\hat\lambda_0)$ to be
non-recursive.)

 Our question about the value of $S$ is in
a sense equivalent to asking how asymptotically close one can minorize
the non-recursive function $s(N)$ by a recursive function. Our conjecture
is that the value of $S$ defined by (1) is strictly less than one. \par
The argument in favour of this conjecture is the following one. Note that the
non-recursiveness of $s(N)$ is a corollary of
the unsolvability of the halting problem for
Turing machines. Moreover, the mentioned proofs of unrecognizability
of $S_0$ and $S^n$ for $n\geq 5$ rely on the
existence of an effective procedure which for a given Turing machine $\tau$ and
its arbitrary input $w$ constructs a triangulation $T$ of a smooth manifold $M$
in such a manner that $\tau$ starting its work from $w$ will eventually halt
if and only if $M$ will be homeomorphic (or PL-homeomorphic,
or diffeomorphic) to
$S_0$ (or, correspondingly, to $S^n$). Thus, it seems reasonable to check
whether or not the recursion-theoretic analogue of the conjecture
$S<1$ will hold. This analogue can be formulated as follows:\par
Let $U$ be a universal Turing machine. Let $H_U(N)$ denote the number of
its inputs $w$ of length $\leq N$ such that $U$ starting to work from $w$
will eventually halt. The unsolvability of the halting problem implies that
$H_U(N)$ will be a non-recursive function. Let $A$ be an algorithm which
answers for any input $w$ of $U$ whether or not $U$ starting
to work from $w$ eventually halts , but which is permitted to say
``do not know" concerning some inputs.
(Such algorithms obviously exist.) Let $H_{UA}(N)$ denote
the number of inputs of length $\leq N$ for which $A$ tells that
$U$ eventually halts. We are interested in $H(U)$ which is defined
as $\sup_A\lim\ \sup_{N\longrightarrow\infty}{H_{UA}(N)\over H_U(N)}$.
Moreover, we are interested in ``natural" universal Turing machines for
which all inputs are meaningful and which have no redundancy. (Examples
of such machines are given in the book [C1].)
For ``natural" Turing machines $H(U)<1$.
The proof of this fact was given by Chaitin ([C3]) using algorithmic
information theory.
(Good expositions of algorithmic information theory can be found in
the book [C1] and reviews [LV] and [ZL]).  Thus, G. Chaitin proved
a recursion-theoretic analogue of our conjecture ``$S<1$". We would like to
mention a paper [S] which also contains a related result in logic but uses a
different approach.\par
\medskip
It should be noted that a possible absence
of computability in Quantum Gravity theories
was mentioned in the Geroch and Hartle paper [GH].
In [GH] it was observed that if one needs to perform a summation
over all possible topological types of four dimensional PL-
manifolds then this can turn out to be impossible due to the fact that
the identification of the topological type of a
given manifold is an unsolvable problem. In this paper we discuss a Markov
process for {\it one fixed} topological type of manifolds. Thus, the
absence of
computability discussed in the present paper is essentially different
from the result of Geroch and Hartle. The possibility of non-computability
arising in Statistical Mechanics models was also discussed in [K].
\par

{\bf Acknowledgements.} We are grateful to Prof. Gregory Chaitin who
explained to us how to prove a recursion-theoretic
analogue of our conjecture ``$S<1$"
using his results in the algorithmic information theory. We wish to thank
Prof. Gregory Mints and Roni Rosner for the fruitful discussions. \par
We wish to thank A. Migdal for introducing us to this field and to
M. Agishtein for interesting discussions.
A significant part of this work was done during the visit of A.Nabutovsky
to Stanford University. He wants to thank the Department of Mathematics of
Stanford University for its kind hospitality.

\vskip 2cm
\underline{REFRENCES}

${\rm [ABB]}$ F. Acquistapace, R. Benedetti and F. Broglia, \
Inventiones Mathematicae  {\bf 102} (1990), 141.

${\rm [AM]}$ M.E. Agishtein, A.A. Migdal, preprint PUPT-1287 (October 1991) and
             PUPT-1311 (March 1992).

${\rm [A]} $ J.W. Alexander, Ann. Math. {\bf 31} (1930), 292.

${\rm [ADF]}$ J. Ambjorn, B. Durhuus, J. Frochlich,\ Nuclear Physics {\bf B270}
             ${\rm [FS16]}$ (1986),457.

${\rm [AJ]}$ J. Ambjorn, J. Jurkiewicz, preprint NBI-HE-91-47

${\rm [BD]}$ A. Billoire, F. David, Nuclear Physics {\bf 275}  ${\rm [FS17]}$
             (1986), 617.

${\rm [BHP]}$ W. Boone, H. Haken and V. Poenaru,
in ``Contributions to Mathematical
             Logic", ed. Arnold Schmidt, Schutte and Thiele, North-Holland,
             1968.

${\rm [BKKM]}$ D.V. Boulatov, V.A. Kazakov, I.K. Kostov, A.A. Migdal, Nuclear
             Physics {\bf B275} ${\rm [FS17]}$ (1986),641.

${\rm [C1]}$ G. Chaitin, "Information, Randomness and Incompleteness",
                         Singapore, Word Scientific, 1987.

${\rm [C2]}$ G. Chaitin, Computers \& Mathematics With Applications, {\bf 2}
             (1976), 233. Also appears as a reprint in [C1].

${\rm [C3]}$ G. Chaitin, personal communication.

${\rm [F]}$  A.T. Fomenko, "Differential Geometry and Topology", Consultants
             Bureau, New York and London, 1987.

${\rm [GH]}$  R. Gerosch and J. Hartle, Foundations of Physics {\bf 16} (1986),
              533.


${\rm [GV]}$ M. Gross, S. Varsted preprint NBI-HE-91-33.


${\rm [K]}$ I. Kanter, Phys. Rev. Lett {\bf 64} (1990), 332.

${\rm [LV]}$ M. Li, P.M.B. Vitanyi "Kolmogorov Complexity and its application",
pp. 187-254 in "Handbook of Theoretical Computer Science", ed. by Jan van
             Leeuwen, Elsevier, 1990.

${\rm [M] }$ A.A. Migdal, J. of Geometry and Physics, {\bf 5} (1988), 711.

${\rm [Po]}$ A.M.Polyakov, Phys. Lett. {\bf B 103} (1981), 207.



${\rm [S]}$ A.O. Slisenko, Zapiski LOMI {\bf 20} (1971), 200 (in Russian).

${\rm [V]}$ S. Varsted, preprint UCSD/PTH 92/03 (January 1992)

${\rm [VKF]}$ I.A. Volodin, V.E. Kuznetzov and A.T. Fomenko, Russian Math.
Surv.
             {\bf 29}(5) (1974), 71-172.

${\rm [ZL]}$ A.K. Zvonkin, L.A. Levin,
             Russian Math. Surv. {\bf 25}(1970), 83-124.
\end{document}